\documentclass[9pt,twocolumn,twoside]{osajnl}

\journal{ol} 

\usepackage{textgreek}
\usepackage[utf8]{inputenc}
\usepackage[T1]{fontenc}
\usepackage{soul}
\usepackage{titlesec}

\setboolean{shortarticle}{true}

\title{All-fiber laser source at 1645~nm for lidar measurement of methane concentration and wind velocity}

\author[1,*]{Philippe Benoit}
\author[1,*]{Simon Le Méhauté}
\author[1,$\dagger$]{Julien Le Gouët}
\author[1,2]{Guillaume Canat}

\affil[1]{O.N.E.R.A. (Office National d'Etudes et Recherches Aérospatiales), 91120 Palaiseau, France}
\affil[*]{at O.N.E.R.A. during this work}
\affil[2]{now at Keopsys Industries, Lumibird, 22300 Lannion, France}

\affil[$\dagger$]{Corresponding author: julien.le\_gouet@onera.fr}

\begin{abstract}
We report on the realization of an all-fiber laser source that delivers single-frequency pulses at 1645~nm, on a linearly polarized single-mode beam, based on stimulated Raman scattering in passive fibers. The pulse energy reaches 14 \textmu J, for a repetition rate of 20~kHz, and the spectral linewidth is 9.5~MHz, almost Fourier-transform limited for the 100~ns square pulses. To the best of our knowledge, this energy is the highest reported at 1645~nm in an all-fiber laser source. Our method consists in reducing the stimulated Brillouin scattering (SBS) gain for the pump and signal pulses, respectively by sweeping the optical frequency of the pump beam, and by applying a strain gradient on the amplification fiber. This compact laser source is now used in a transportable lidar system to measure simultaneously wind velocity and methane (CH$_4$) concentration.
\end{abstract}

\setboolean{displaycopyright}{true}

\begin{document}

\maketitle

Laser sources at 1645~nm are of high interest for remote methane sensing in transportable systems \cite{Riris_2012}. In particular, our long-term goal is to provide a lidar measurement of methane flux, with a distance resolution. For this we aim simultaneous measurements of methane concentration, by differential absorption lidar (DIAL) technique, together with the velocity of the aerosols that scatter the light back to the system, by heterodyne analysis of the Doppler shift. The distance resolution can be provided by a time-of-flight measurement. 

The corresponding laser source must deliver square pulses at wavelengths close to 1645~nm \cite{Cezard_2016}, with a spectral linewidth close to the Fourier-transform (FT) limit to preserve the velocity resolution, and a single-spatial mode to optimize pointing precision and coherent mixing efficiency. The pulse energy must be the highest possible, as it determines the maximum range of the lidar measurement.

Fiber-based amplifiers are attractive for their compactness and robustness to harsh environments. For laser wavelengths up to about 2.1~\textmu m, the most efficient way to amplify a signal beam in a silica fiber source is to use stimulated emission from a well-known rare-earth ion (ytterbium, erbium, thulium...). However, the 1.6-1.7~\textmu m band cannot be covered by stimulated emission to a ground level of such ions in a silica fiber.

Thanks to the well-known progress accomplished in the field of optical fiber telecommunications, the power delivered by single-frequency fiber laser sources increased tremendously. In particular, in erbium-doped fibers, the spectral band close to 1.55~\textmu m offers the highest gain. By chance, the Stokes spectral shift experienced by an intense laser field through stimulated Raman scattering (SRS) is about 400~cm$^{-1}$ in silica fibers \cite{Stolen_1973}, which corresponds to a wavelength increase of 110~nm. Therefore, the 2$\nu_3$ band of methane (between 1640~nm and 1690~nm) becomes accessible through Raman amplification, by pumping a silica fiber with an erbium-doped fiber laser \cite{Horiguchi_1992}.

With this method, a major obstacle resides in the stimulated Brillouin scattering (SBS), which can degrade the transmission of narrow-linewidth signals. Indeed, the SBS gain in a silica fiber is higher than for SRS by two orders of magnitude, and so is the corresponding critical power \cite{Agrawal_2013}. Therefore, reducing the SBS gain represents a high stake to improve the efficiency of a single-frequency Raman fiber amplifier (RFA) \cite{Zhang_2012, Dajani_2013}.


The SBS effect arises from the phase-matching between a high intensity optical wave and the acoustic waves that it generates in the propagation medium. Reducing the SBS gain can thus consist in reducing the optical intensity, or breaking the phase-matching along the fiber. The first approach is not compatible with an efficient Raman amplification, which also requires a high intensity. The second approach can be materialized by reducing the lifetime of the acoustic phonons \cite{Gray_2009}, or by broadening inhomogeneously the SBS gain. This last method consists in introducing a longitudinal variation of the acoustic velocity, related to the glass density, through thermal \cite{Hansryd_2001,Dajani_2013} or strain \cite{Yoshizawa_1993, Zhang_2012} gradients. In the particular case of a Raman amplifier, it is also possible to broaden the Brillouin Stokes wave through cross-phase modulation, by modulating the pump intensity \cite{Harish_2019}.

To generate and amplify our pulsed signal we use high power pump pulses, which could also be affected by SBS. However the SRS gain spectrum (several THz) is much wider than the SBS gain (about 40~MHz) \cite{Agrawal_2013}, so the linewidth of the RFA pump can be largely broadened to avoid SBS on this wave \cite{Lichtman_1987}. 

\begin{figure*}[htbp]
\centering
\includegraphics[width=.85\linewidth, trim={0 .2cm 0 0}, clip]{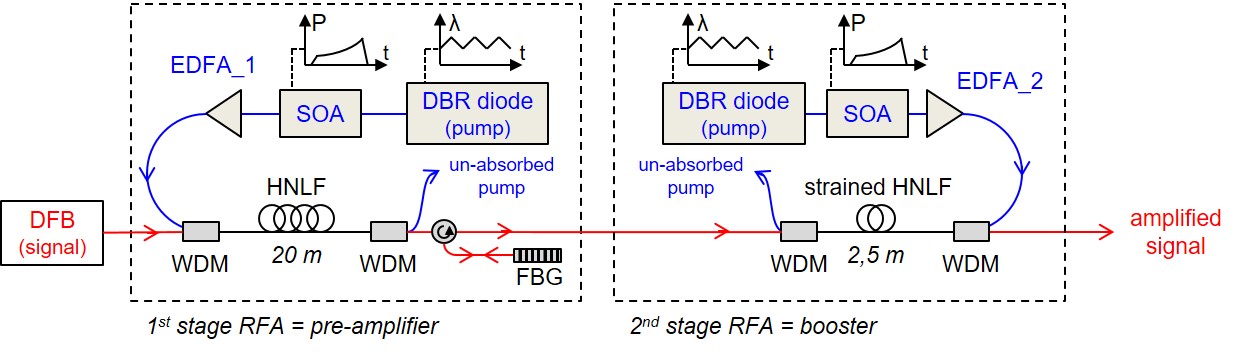}
\caption{Setup of the double-stage Raman amplifier. The temporal aspects are indicated for the optical power of the semiconductor optical amplifiers (SOA), and for the wavelength of the DBR diodes. The HNLF lengths are justified in the main text.}
\label{fig:experimental_setup}
\end{figure*}

In this article we describe an all-fiber laser source of single-frequency pulses at 1645~nm, for simultaneous measurements of methane concentration and wind velocity. 
With the methods detailed below we demonstrate the highest peak power reported so far, to the best of our knowledge, for a single-frequency all-fiber laser in this spectral domain. This source has already been used on field, with longer pulses, to monitor methane concentration with a resolution better than 20~ppm at 150 m, with a range resolution of 30~m and an integration time of 10~s \cite{Cezard_2020}.



The configuration of our master oscillator power amplifier (MOPA) is presented in Figure \ref{fig:experimental_setup}. The seed is a continuous wave (cw) linearly polarized distributed Bragg reflector (DBR) laser diode from Toptica, emitting 20~mW at $\lambda_\text{sig}$=1645.5~nm with a 0.96~MHz spectral linewidth. For heterodyne Doppler measurements of wind velocity, a fraction of the signal field must be used as a local oscillator, so the power injected in the amplifier is limited to about 1~mW.

The cw seed is converted into a pulsed signal by switching the Raman pump beams, and the signal pulses are amplified in two successive fiber amplifiers. In both amplifiers, the Raman pump source is a MOPA itself, constituted of a cw tunable narrow linewidth seed laser ($\Delta\nu \simeq 2$~MHz) emitting at $\lambda_\text{pump}$=1545.3~nm, a semiconductor optical amplifier (SOA) to switch and shape the pulses, and an erbium-doped fiber amplifier (EDFA). For our lidar application, we choose a pulse duration $\tau_p=100$~ns and a repetition rate $f_\text{rep}=20$~kHz. 

The same highly nonlinear fibers (HNLF) is used in the two amplification stages (CorActive SCF-UN-3/125-25-PM). It has a mode field diameter MFD=4.7 \textmu m at $\lambda_\text{pump}$, and is normally dispersive at both $\lambda_\text{pump}$ and $\lambda_\text{sig}$ to avoid modulation instability for the pump and the signal. According to our own pump-probe measurements in small-signal regime \cite{Stolen_1973}, the maximum SRS gain is $g_R=8.0 \times 10^{-14}$~m/W at $\lambda_\text{pump}$, for a 13~THz spectral shift between the pump and the signal. Similarly, for the SBS gain, we measured $g_B=1.3 \times 10^{-11}$~m/W. The propagation loss is about 1.3~dB/km, according to the manufacturer. Like all the other fiber components of the setup, the HNLF has a polarization-maintaining structure.

In the first stage (pre-amplifier), the pump and signal beams are coupled into the HNLF with a wavelength division multiplexer (WDM), and separated at the output by another WDM. Pump and signal are co-propagative to maximize the pump-to-signal conversion efficiency. Indeed, counter-propagating pump pulses would have to be longer (with the same peak power) than the signal pulse to guarantee the overlap over all the interaction length. In order to remove the residual pump at the output of this stage, as well as the Raman Amplified Spontaneous Emission, we insert a circulator and a reflective fiber Bragg grating (FBG, R=99\%) centered at $\lambda_\text{sig}$, with a 3dB linewidth of 1~nm. This arrangement also blocks any Stokes wave from the second stage that could seed SBS in the first stage.

The peak power of the pump pulses is a crucial parameter to evaluate the maximum signal energy that can be extracted from the Raman amplifiers. While our signal has to be single-frequency and is exposed to SBS, the Raman pump may be spectrally wide, because the Raman amplification does not rely on a phase-matching condition between pump and signal. In particular, the pump spectrum can be much wider than the SBS gain linewidth $\Delta\nu_\text{SBS}$ (typically 40~MHz at 1.5 \textmu m in a silica fiber). Here we spectrally broaden the Raman pump fields, in both stages, by applying electronic modulations on the refractive index of the DBR of each seed laser (see \ref{fig:experimental_setup}).

The principle of SBS gain reduction by using a linear frequency chirp is well explained \cite{Mungan_2012}, and has already been implemented in a Raman fiber amplifier for a short range measurement of methane concentration \cite{Mitchell_2009}. For a chirp rate $\beta$, the spectral broadening of the pump writes as:
\begin{equation}
\Delta\nu_\text{pump} = \Delta\nu_\text{SBS} + \beta\tau_\text{int}
\end{equation}
where $\tau_\text{int}$ is the duration of the interaction of the pump pulse with the fiber of length $L_f$ and core refractive index $n$. If the pulse is longer than the transit time $\tau_f=nL_f/c$, then $\tau_\text{int}=\tau_f$. Otherwise $\tau_\text{int}=\tau_p$. 

\begin{figure}[htbp]
\centering
\centering\includegraphics[width=.75\linewidth, trim={0 5.5cm 6cm 0}, clip]{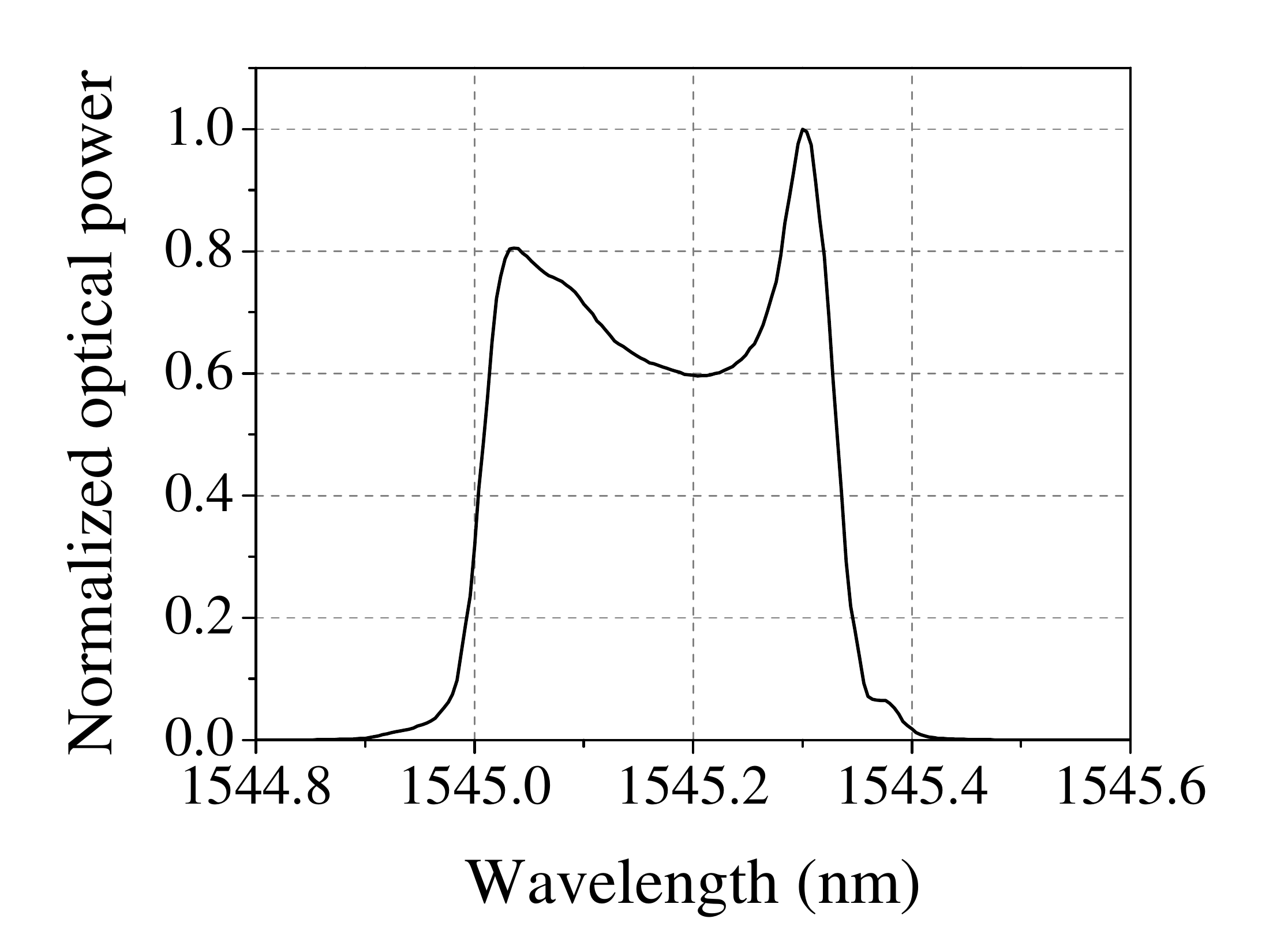} 
\caption{Optical spectrum of the chirped laser diodes that seed the Raman pump beams in both stages. The resolution of the optical spectrum analyzer is 0.05~nm.}
\label{fig:output_finisar_spectrum}
\end{figure}

Ultimately, the amplitude of the frequency ramp is limited by the free spectral range (FSR) of the seed laser: FSR$\simeq 40$~GHz ($\Delta\lambda \simeq 0.33$~nm) for our laser diodes (Finisar S7500).
As illustrated on Figure \ref{fig:output_finisar_spectrum}, the full FSR of the lasers can be scanned by a triangular modulation with a repetition rate as fast as $f_\text{mod}=40$~MHz. Beyond this value, the frequency scan is not linear anymore. The ramp duration $\tau_\text{ramp}=1/(2f_\text{mod})=12.5$~ns is thus the shortest time during which the frequency can be swept across the entire FSR. Then as long as the fiber lengths are longer than $L_f=c/(2nf_\text{mod})=2.50$~m, the SBS gain is reduced by a factor $1 + \text{FSR}/\Delta\nu_\text{SBS} \simeq 1000$. In reality the SBS gain reduction is slightly lower than 30~dB, because the frequency modulation is not perfectly triangular, as illustrated by the non-flat spectrum (Fig. \ref{fig:output_finisar_spectrum}). However we find that the pump power is not limited by SBS in the pre-amplifier or in the second stage.

We can now determine the optimal length of the first stage HNLF by calculating the output signal power and the SBS critical power, for various fiber length values. The optimal fiber length corresponds to the value for which the two powers are equal. For the pump amplifier EDFA\_1 (realized by Lumibird), the average power can reach $\langle P_{p1} \rangle=400$~mW. For $\tau_p=100$~ns and $f_\text{rep}=20$~kHz, the corresponding peak power is $P_{p1}^\text{peak}=200$~W. Given the mode area $A$ in the HNLF, the fiber length should thus be shorter than $L_{max}=21A/(P_{p1}g_B)=0.14$~m to avoid SBS. Thanks to the 30~dB reduction of the SBS gain by frequency chirp of the pump seed, the fiber length could extend to more than 100~m. Finally we calculate that the optimal HNLF length is $L_1=20$~m. With this length, we measure a signal peak power $P_\text{sig}^\text{peak}=3$~W at the pre-amplifier output.

In the second Raman amplification stage (booster), the pump is counter-propagative compared to the signal, so as to attenuate the transfer of any power fluctuation of the pump field towards the signal \cite{Fludger_2001, Liu_2018}. In particular, this configuration helps maintaining the Raman gain almost constant during the signal pulse, despite the pump power modulation that is described below.

As for the pump pulse duration in the counter-propagative configuration, it must be a little longer than the signal in order to guarantee the temporal overlap and a constant Raman gain during the signal pulse. More precisely, the pump pulse should arrive already at the fiber input when the signal enters, and leave the fiber output when the signal comes out. For a signal duration $\tau_\text{sig}$ and a fiber length $L_2$, this implies a pump duration $\tau_\text{pump}=\tau_\text{sig} + 2nL_2/c$.

The available average power for the pump amplifier EDFA\_2 of the booster is limited to $\langle P_{p2}\rangle \simeq 2$~W, which corresponds to a peak power $P_{p2}^\text{peak}=1$~kW. The fiber length should be shorter than $L_\text{max}=21A/(P_{p2}^\text{peak}g_B)\simeq 30$~mm to avoid SBS, but with the same frequency chirp presented for the pre-amplifier, the fiber length can be extended to 30~m. 

However in the booster, the peak power of the single-frequency signal can reach the SBS critical power. In order to alleviate this limitation, we apply a longitudinal strain gradient along the HNLF, with a triangular shape that guarantees a minimal tension at both ends of the fiber \cite{Canat_2014,Lucas_2014}. The differences of silica density and corresponding sound velocity between the different sections of the fiber result in a inhomogeneous broadening of the SBS gain spectrum. Hence the phase-matching between the signal, Stokes and local acoustic waves is limited to the small sections of identical sound velocity, and the effective length of the SBS interaction is reduced. In practice, we can obtain a flat SBS gain with a spectral width of 250~MHz, which corresponds to a relative extension of 0.5\% of the fiber length \cite{Boggio_2005}. Broadening the SBS gain spectrum from 40~MHz to 250~MHz reduces $g_B$ by a factor of 6. Considering this factor, we finally calculate the optimal HNLF length for the booster, and find $L_2=2.5$~m.



\begin{figure}[htbp!]
\centering
\centering\includegraphics[width=.7\linewidth, trim={0 5cm 5cm 0}, clip]{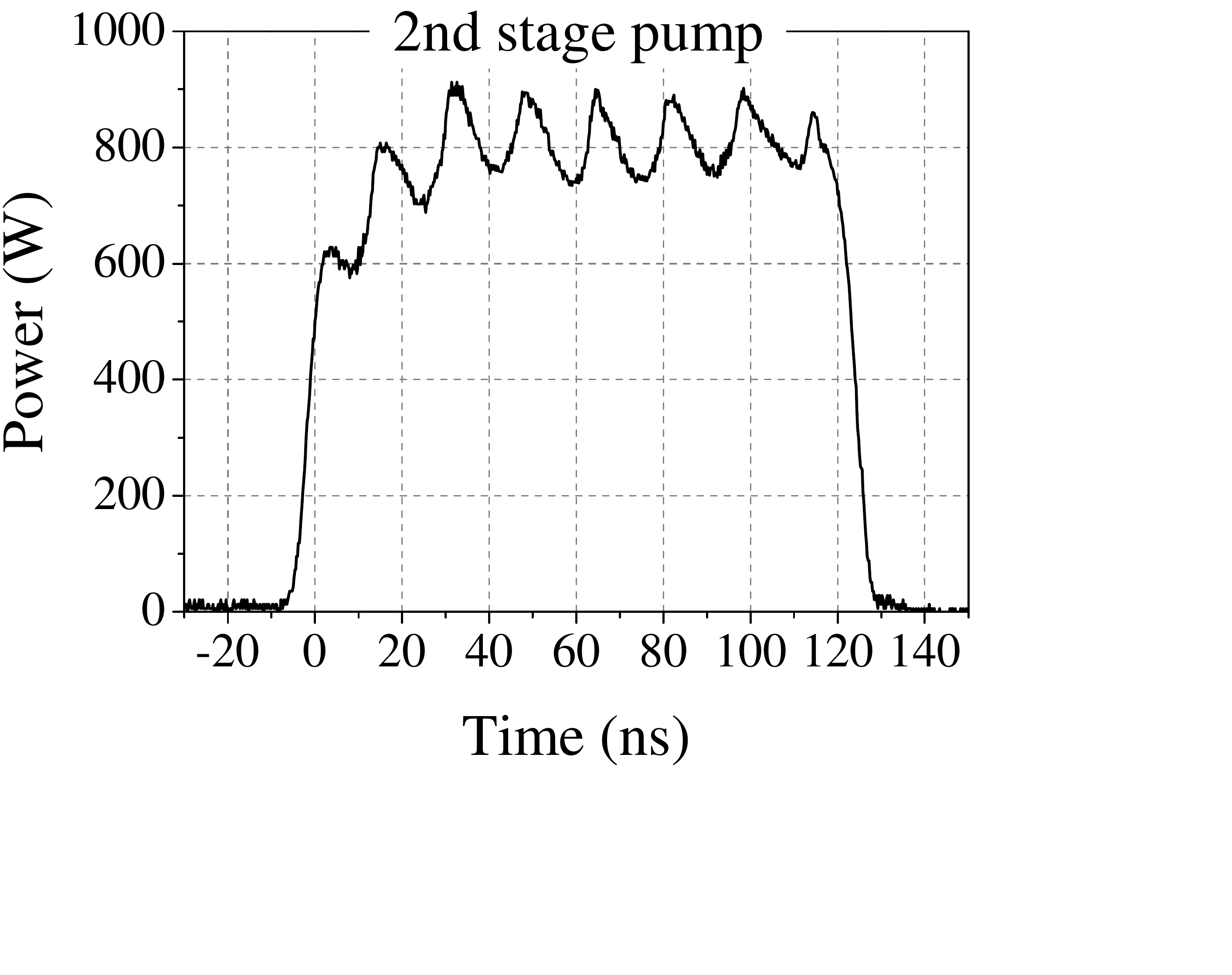}
\caption{Power of the second stage Raman pump pulses, as measured at the output of EDFA\_2.}
\label{fig:output_edfa}
\end{figure}

Concerning the shape of the pump pulses, they must be as flat as possible to avoid Raman gain variations during the signal pulse, which would translate into phase variations and spectral broadening \cite{Benoit_2020}. At the output of EDFA\_1 and EDFA\_2, two main effects must be taken into account. First, it is well-known that the amplification of laser pulses in saturated regime leads to a shape distorsion, due to higher available gain at the beginning of the pulse than at the end \cite{Frantz_Nodvik}. A standard method consists in shaping the input pulse to pre-compensate the distorsion \cite{Schimpf_2008}. Here the pulse shaping is realized in each pump source by adjusting the pattern of the SOA transmission (see Fig. \ref{fig:experimental_setup}). The result of the compensation at the output of the pump source of the booster (EDFA\_2) is shown on Figure \ref{fig:output_edfa}, which illustrates a second possible cause of gain variation. We apply the same compensation method on the pulse shape of the pre-amplifier pump (EDFA\_1), which can also be used to correct any residual distortion in the booster.

Indeed, on top of the square envelope, the pump pulses at the output of EDFA\_2 exhibit a power modulation with a relative amplitude of 20\% peak-to-peak. The power modulation is synchronous with the triangular frequency modulation of the DBR laser diode. It also appears on the laser output signal, but with a relative amplitude ten times smaller. Therefore it seems that the conversion from frequency to power modulation occurs mainly in the EDFA. Since there is no spectral filtering in the pump sources, one explanation could be a spatial filtering between the slightly multimode large mode area (LMA) fibers at the EDFAs outputs, and the single mode fiber of the WDM. The same effect was observed in a homemade EDFA, also based on a LMA fiber.

\begin{figure}[htbp!]
\centering\includegraphics[width=.8\linewidth, trim={0 0 -2.cm 0}, clip]{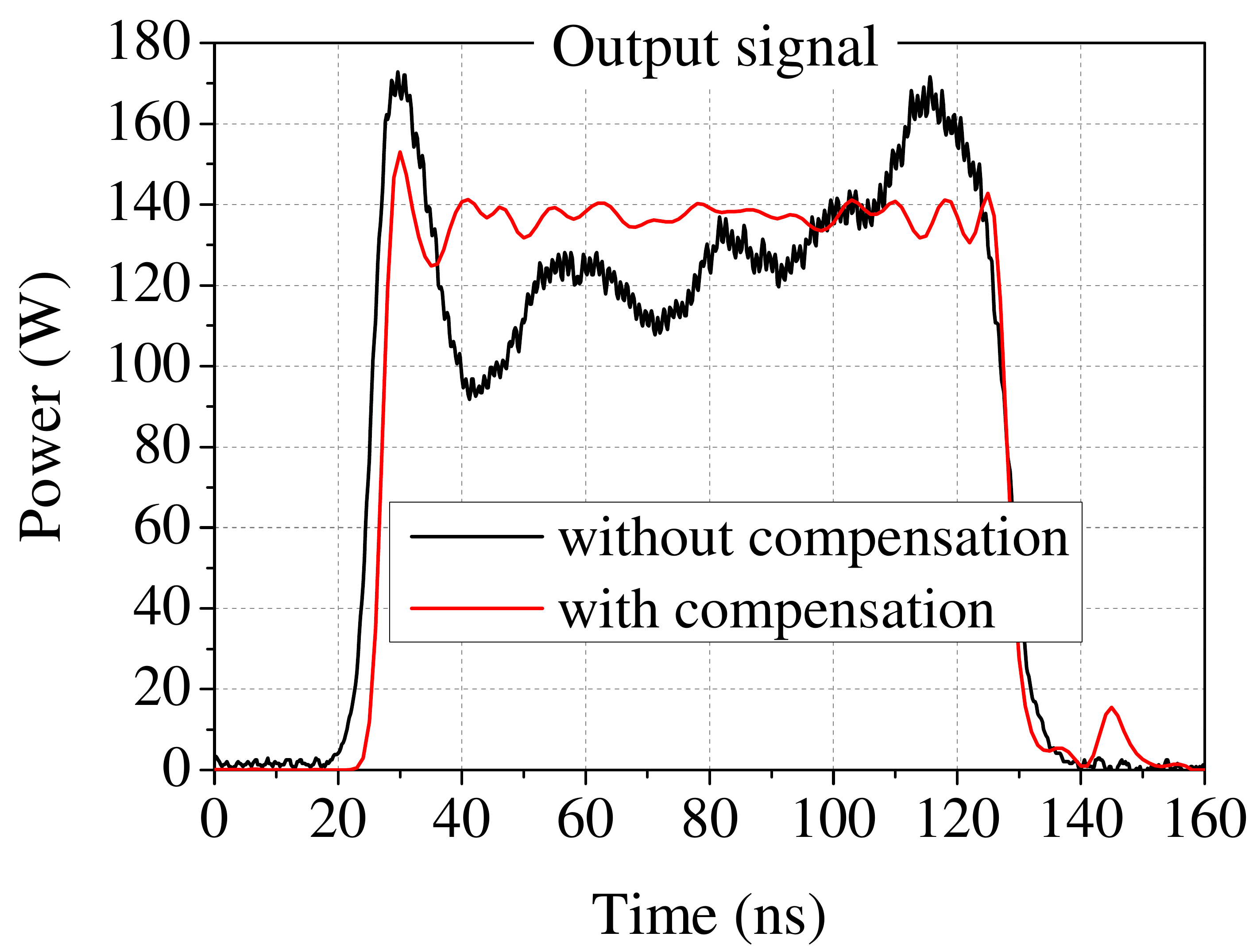} 
\caption{Power of the signal pulses at the RFA output, before and after (respectively black and red curves) correction of the pump modulation. The compensation is obtained by tuning the transmission of the SOA of the pre-amplifier pump.}
\label{fig:Raman_output_time}
\end{figure}

Due to the short lifetime of the optical phonons that drive the SRS, such variations of the pump power will directly translate in variations of the gain and of the amplified signal. Figure \ref{fig:Raman_output_time} gives an example of the signal shape that results from the power modulation of the pump, before correction (black curve). In the Raman pre-amplifier, we compensate the pump modulation by adjusting the transmission of the SOA in the pump source.


In the booster, the SOA dynamic would not be sufficient to fully compensate the modulation. We thus chose the counter-propagative configuration of pump and signal beams, as it averages the variations of the pump power, which would transfer to the signal power \cite{Fludger_2001}. The corrected signal shape is shown on Figure \ref{fig:Raman_output_time} (red curve). We find that the signal peak power at the RFA output reaches $P_\text{sig}^\text{peak}=140$~W for $\tau_p=100$~ns and $f_\text{rep}=20$~kHz, which corresponds to a pulse energy $E_s=14$~\textmu J. This maximum signal power is limited by SBS.





\begin{figure}[htbp]
\centering\includegraphics[width=.75\linewidth, trim={0 4.5cm 5cm 0}, clip]{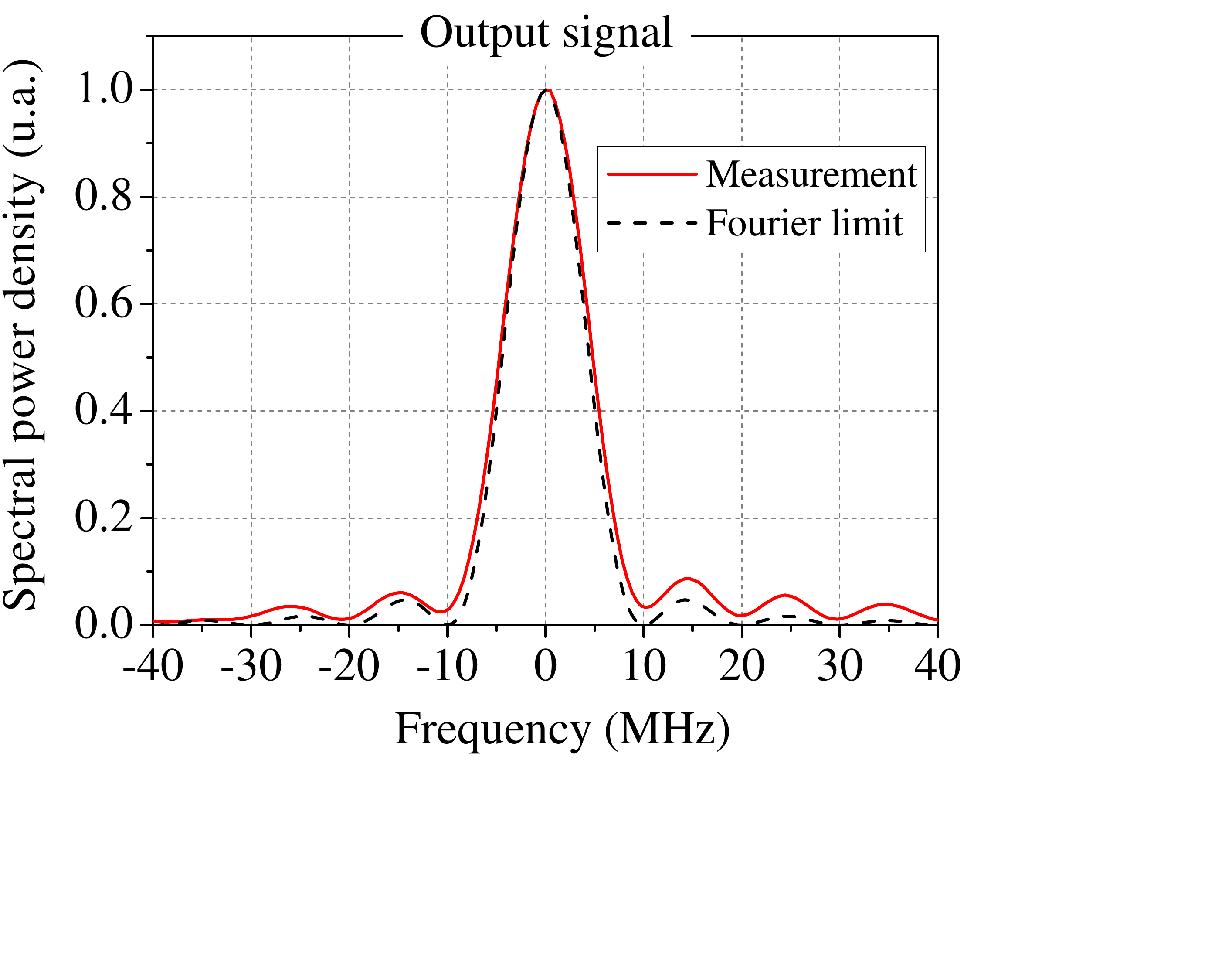} 
\caption{Optical spectrum of the amplified signal at the output of the RFA second stage and the FT limit for a 100~ns square pulse, measured with an heterodyne setup.}
\label{fig:Raman_output_spectral}
\end{figure}

Finally we measure the frequency linewidth of the output signal from the heterodyne interference between the signal and a frequency shifted fraction of the seed laser \cite{Okoshi_1980}. Figure \ref{fig:Raman_output_spectral} presents the spectrum of the output signal, together with the spectrum of the FT of an ideal 100~ns square pulse. The full width at half-maximum is 9.5~MHz, so the time-bandwidth product (TBP) is 0.95. For a square pulse the TBP is 0.89, so the signal linewidth is only larger by 7\% compared to the FT-limited spectrum. For the coherent lidar application, this translates into preserved sensitivity for wind speed measurement.

The energy outside the central lobe amounts to 20\%, instead of 10\% for the ideal FT spectrum of the square pulse. We attribute this increase to the residual power modulation on the output signal (Figure \ref{fig:Raman_output_time}). Self-phase modulation (SPM) is not expected to be significant, since the optical intensity is almost constant over the signal pulses. Only the edges can thus experience broadening,but their contribution to the total energy is negligible. Moreover, SPM is proportional to the average of the signal intensity across the fiber, which remains low thanks to the exponential growth of the signal.



To conclude, we demonstrate the generation of single-frequency 14~\textmu J pulses at 1645~nm in an all-fiber source based on stimulated Raman amplification. To the best of our knowledge, this corresponds to the highest energy reported at this wavelength for a single-frequency source, on a linearly polarized single-mode fiber. For the 100~ns pulse duration, the linewidth is 9.5~MHz (close to the FT limit) and the peak power is about 140~W. To avoid SBS limitation on the pump power, we apply a triangular modulation of the pumps optical frequency. The modulation increases the SBS critical power by almost 3 orders of magnitude. As for the signal beam, we increase its SBS critical power by a factor of 6, by applying a strain gradient on the booster stage fiber.

The laser source is presently used in a transportable DIAL system for simultaneous remote measurements of wind velocity and CH$_4$ concentration. Obviously the long term perspective consists in raising the lidar range by increasing the pulse energy. Several approaches can be explored, by increasing the pump power, and by working on the HNLF itself to increase the ratio $g_R/g_B$ between Raman and Brillouin gains.

\section*{F\lowercase{unding}. \hfill \normalfont {T\lowercase{his work was funded by the} NAOMI \lowercase{project} (ONERA-T\lowercase{otal}).}}

\section*{A\lowercase{cknowledgment}. \normalfont T\lowercase{he authors are grateful to} N. C\lowercase{ézard}, L. L\lowercase{ombard}, A. D\lowercase{urécu}, F. G\lowercase{ustave, and} A. M\lowercase{ussot for fruitful discussions.}}

\section*{D\lowercase{isclosures}. \normalfont T\lowercase{he authors declare no conflicts of interest.} }

\bibliography{2020_ampli_1645nm_bib}

\bibliographyfullrefs{2020_ampli_1645nm_bib}

\end{document}